\documentclass{aa}

\usepackage{hyperref}
\usepackage{mathrsfs}
\usepackage{threeparttable}

\begin{document}

\title{Line-force driven wind from a thin disk in tidal disruption event}
\titlerunning{Winds from TDEs}

\author{De-Fu Bu\inst{1}\thanks{Corresponding author: De-Fu Bu}
        \and Xiao-Hong Yang\inst{2}\thanks{Corresponding author: Xiao-Hong Yang}
        \and Liang Chen\inst{3}
        \and Chenwei Yang\inst{4}
        \and Guobin Mou\inst{5}
        }
\authorrunning{Bu, Yang, Chen, Yang, Mou}

\institute{Shanghai Key Lab for Astrophysics, Shanghai Normal University, 100 Guilin Road, Shanghai 200234, China\\
\email{dfbu@shnu.edu.cn}
\and Department of Physics and Chongqing Key Laboratory for Strongly Coupled Physics, Chongqing University, Chongqing 400044, People's Republic of China\\
            \email{yangxh@cqu.edu.cn}
            \and Shanghai Astronomical Observatory, Chinese Academy of Sciences, 80 Nandan Road, Shanghai 200030, China\\
            \and Polar Research Institute of China, 451, Jinqiao Road, Pudong, Shanghai, 200136, China\\
            \and Department of Physics, Nanjing Normal University, Nanjing 210023, China}

\abstract
{}
   {Winds from the accretion disk in tidal disruption events (TDEs) play a key role in determining the radiation of TDEs. The winds from the super-Eddington accretion phase in TDEs have recently been studied. However, properties of the winds from the sub-Eddington accretion disk in TDEs are not clear. We aim to investigate properties of winds from the circularized sub-Eddington accretion disk in TDEs. We study the line force driven accretion disk wind.}
   {We perform two-dimensional hydrodynamic simulations using the PLUTO code to study the line force driven wind from the circularized accretion disk around a $10^6$ solar mass black hole in TDEs.}
   {We find that although the disk has a very small size in TDEs, strong wind can be driven by line force when the disk have luminosity higher than $20\%$ of the Eddington luminosity. The maximum velocity of wind can be as high as $0.3$ times the speed of light. The kinematic power of wind is in the range of $1\%-6\%$ times the Eddington luminosity. Strong wind can be driven by line force from the thin disk around a $10^6$ solar mass black hole in TDEs. We briefly discuss the possible radio emission from the shock when the wind collides with the surrounding medium.}
   {}
\keywords{accretion, accretion disks -- black hole physics -- quasars: supermassive black holes.}
\maketitle

\section{Introduction} \label{sec:intro}

In a galaxy, some stars can occasionally move to the vicinity of the central super-massive black hole (SMBH) at the galaxy center. The tidal radius of the black hole is defined as $R_{\rm{T}} = (M_{\rm BH}/M_*)^{1/3}R_*$, with $M_{\rm BH}$, $M_*$ and $R_*$ being the black hole mass, the mass of the star and the radius of the star, respectively. Whether the star will be disrupted depends on the penetration factor $\beta$, which is defined as $\beta = R_{\rm T}/R_{\rm p}$, with $R_{\rm p}$ being the stellar orbital pericenter. The so-called tidal disruption event (TDE; \citep{Rees1988,Evans1989}) is triggered if the pericenter of the stellar orbit ($R_{\rm p}$) is smaller than the tidal radius $R_{\rm{T}}$ ($\beta \ge 1$). For a parabolic stellar orbit, roughly half of the disrupted debris is bound and falls back to the pericenter. The fallback time of the bound debris is $t_{\rm fb} \approx 41 (M_{\rm BH}/10^6M_\odot)^{1/2}(M_*/M_\odot)^{-1}(R_*/R_\odot)^{3/2} \beta^{-3}$ days, with $M_\odot$ and $R_\odot$ being solar mass and solar radius, respectively \citep{Lodato2011}. If the specific energy distribution is flat within the disrupted star, the debris fallback rate $\dot M_{\rm fb} = 1/3(M_*/t_{\rm fb})(t/t_{\rm fb})^{-5/3}$ \citep{Rees1988, Phinney1989}. For a $10^6M_\odot$ black hole disrupting a solar type star, the peak fallback rate can be as high as 133 Eddington accretion rate ($\dot M_{\rm Edd}$) with $\beta=1$. The debris fallback rate transients from the super-Eddington regime to sub-Eddington regime at a time $t_{\rm cr} = 760 (\eta/0.1)^{0.6}(M_{\rm BH}/10^6M_\odot)^{-2/5}(M_*/M_\odot)^{1/5}(R_*/R_\odot)^{3/5} \beta^{-6/5}$ days \citep{Lodato2011}, with $\eta$ being the radiative efficiency of the accretion process. Some portion of the fallback debris can be accreted by the black hole, which produces a transient luminous electromagnetic radiation.

The bound debris is predicted to fallback with a rate declining with time as $\dot M_{\rm fb} \propto t^{-5/3}$ \citep{Rees1988}. The TDEs first found in soft X-ray bands by \textit{ROSAT} X-ray All-Sky Survey (see \citet{Komossa2015} for review) do have their X-ray light curves declining as $t^{-5/3}$. However, we note that the perfect consistency in the time evolution between the debris fallback rate and X-ray light curve in those TDEs may just be a coincidence. One can not directly predict the TDEs light curve according to the debris fallback rate. The reasons are as follows. First, at early times, the debris fallback rate can deviate from the $t^{-5/3}$ law due to stellar density \citep{Lodato2009}, stellar rotation\citep{Golightly2019} and the eccentricity of the stellar orbit\citep{Hayasaki2013,Hayasaki2018,Park2020,Cufari2022,Zhong2023}. The fallback debris stream collides due to relativistic effects which will result in the formation of an accretion disc \citep{Hayasaki2013,Hayasaki2016,Bonnerot2016}. Second, once an accretion disc forms, the black hole accretion rate is determined by the viscous process, which would also deviate from the $t^{-5/3}$ law. For example, \cite{Cannizzo1990} found that at late time of a TDE, in the absence of wind, the black hole accretion rate declines as $t^{-19/16} $ for zero viscous stress at the innermost stable circular orbit (ISCO) . \cite{Tamilan2024} and \cite{Tamilan2025a} found that in the presence of wind, the mass accretion rate is steeper than the $t^{-19/16}$ law. In the presence of a strong poloidal magnetic fied, \cite{Tamilan2025b} found the accretion rate scales as $t^{-5/2}$. Third, given the black hole accretion rate, the radiative efficiency of the accretion flow is needed to calculate the radiation luminosity. The radiation efficiency should be a function of accretion rate in the initial high super-Eddington accretion rate phase \citep{Ohsuga2005}.

The wind in TDEs is believed to play a key role in solving the observational puzzles. For optical/UV TDEs, the origin of their emission is still under debate. Theoretically, the predicted size of the accretion disk is a few times $10^{13}$ cm if one assumes a solar type star disrupted by a black hole with $10^{6-7} M_\odot$, with $M_\odot$ being solar mass. However, observationally, the inferred optical/UV radiation scale is $10^{14-16}$ cm \citep{Hung2017,Velzen2020,Gezari2021}, which is orders of magnitude larger than the theoretically predicted disk size. In the `reprocessing' model, optically thick wind is launched in the debris accretion process. The soft X-rays emitted in vicinity of the black hole are reprocessed and re-emitted at large scale at optical/UV band \citep{Strubbe2009,Lodato2011,Metzger2016,Roth2016,Dai2018,Curd2019,Uno2020,Piro2020,Bu2022,Parkinson2022,Mageshwaran2023}.  \footnote{Other models explain the optical/UV emission of TDEs include the shock model \citep{Piran2015,Jiang2016,Steinberg2024,Huang2024,Guo2025} and elliptical accretion model \citep{Liu2017,Liu2021,Wevers2022}.}

In addition to optical/UV emission, radio emission is also detected in TDEs. Some of the detected radio emission is from non-jetted TDEs. For example, in the non-jetted TDEs AT2019dsg and ASASSN-14li, radio emission is detected after several tens of days after disruption \citep{Cendes2021,Alexander2016}. The radio emission in these TDEs may be due to the unbound debirs-CNM (circumnuclear medium)  \citep{Guillochon2016,Krolik2016,Yalinewich2019} or disk wind-CNM  \citep{Alexander2016,Hayasaki2023} interactions.
Very recently, it is found that delayed radio emission of TDEs \citep{Alexander2020,Horesh2021,Cendes2022,Perlman2022,Goodwin2022,Sfaradi2022,Zhang2024} may be due to the interaction of TDE wind with the CNM ( \citep{Barniol2013,Matsumoto2021,Matsumoto2024,Zhou2024,Cendes2024}). In addition, the TDE winds can collide with dense clouds surrounding the black hole. The collision produces bow shocks. The power-law electrons can be accelerated in the bow shocks. Recently, theoretical models of wind-cloud interaction find that the radio emission of some TDEs can be well explained \citep{Mou2022,Bu2023a,Lei2024,Zhuang2025}. TDEs jet is also proposed as an alternative explanation for the delayed radio emissions \citep{Teboul2023,Matsumoto2023,Sfaradi2024}.

The TDE winds play important roles not only in regulating TDEs radiation via `reprocessing' process but also in detecting the properties of the surrounding environment of the black hole via wind-induced radio radiation. TDE winds have been studied in the context of `circularized' debris accretion flow. The winds at a snapshot near the peak fallback rate have been studied \citep{Dai2018,Curd2019}. The TDE accretion flow has no steady-state due to the fact that the gas supply rate (or debris fallback rate) to the flow declines with time. Therefore, it is expected that the properties of wind vary with time. In \citet{Thomsen2022}, several discrete simulations with different accretion rates are performed to investigate the time evolution of TDE winds. \citet{Bu2023b} performed hydrodynamical simulations with radiative transfer to study the wind from `circularized' accretion flow in TDEs. Special conditions in TDEs are taken into account in \citet{Bu2023b}. First, gas is injected in the simulations at two times the pericenter of the disrupted star orbit. This location is the theoretically predicted outer boundary of the accretion flow assuming angular momentum conservation of the debris from a parabolic orbit disrupted star. Second, in order to mimic the gas supply to accretion flow by fallback debris, the gas injection rate is set to be the debris fallback rate which declines as $(t/t_{\rm fb})^{-5/3}$ with $t_{\rm fb}$ being the orbital period of the most bound debris. All the above mentioned work focus on the super-Eddington accretion phase in TDEs. It is found that strong wind can be launched by radiation pressure. The speed of wind can be much higher than $0.1 \rm c$, with $\rm c$ being speed of light. \citet{Bu2023b} found that for a solar type star be disrupted, in the circularized super-Eddington accretion phase, a significant fraction of the fallback debris will be lost in radiation pressure driven wind. For a $10^6M_\odot$ black hole, $57\%$ of the fallback debris becomes wind, while for a $10^7M_\odot$ black hole, the value is $85\%$. \citet{Curd2023} perform simulations studying an evolving TDE accretion flow, in which gas supply rate declines as $(t/t_{\rm fb})^{-5/3}$. However, we note that the debris fallback timescale $t_{\rm fb}$ in \citet{Curd2023} is set to an unrealistic shorter value. Therefore, it is not clear to what extent the properties of wind obtained in that work resemble the real case. Winds from the stream-stream collision process are also studied \citep{Jiang2016,Lu2020}.

The debris fallback rate drops to sub-Eddington value after a time period ($t_{\rm cr}$) since disruption. For a $10^6M_\odot$ black hole disrupting a solar type star, $t_{\rm cr} = 760 $ days. In the sub-Eddington accretion phase ($0.01 \dot M_{\rm Edd} \leq \dot M \leq \dot M_{\rm Edd}$, with $\dot M_{\rm Edd}$ being Eddington accretion rate), the accretion disk is expected to be a thin disk. The  detailed properties of wind from the sub-Eddington thin disk in TDEs are poorly known. The accretion disk in luminous active galactic nuclei (AGNs) is believed to be thin disk. There are many analytical and simulation works studying winds from AGNs. However, we note that those results can not be directly applied to TDEs. The reason is that the thin disk in TDEs is quite different from that in AGNs. For example, the size of accretion disk in TDEs is much smaller than that in AGNs. In addition, the thin disk in TDEs has no quasi-steady state due to the fact that the gas supply rate to the disk declines with time as $t^{-5/3}$ law.

There are two main mechanisms driving wind from a thin disk, namely magnetic driven model and radiation line force driven model. The magnetic driven wind model has always being a hot topic and being studied extensively \citep{Blandford1982,Lynden-Bell1996,Li2014,Fukumura2015,Li2022,Wang2022}. Very recently, \cite{Tamilan2024,Tamilan2025a,Tamilan2025b} find that the presence of magnetic driven wind and magnetic field can have strong effects on the time evolution of the mass accretion rate in TDEs.

Gas in thin disk around super-massive black holes is partially ionized. The UV photons from the disk can be absorbed by gas, the cross section of line resonance can be orders of magnitude higher than that of Compton scattering. Therefore, radiation pressure on resonance lines (hereafter line force), can effectively launch wind even from sub-Eddington luminosity system. The line force driven wind have been studied by both analytical work \citep{Murray1995} and numerical simulations \citep{Proga2000,Proga2004,Nomura2016,Nomura2017,Mizumoto2021,Yang2021a,Yang2021b}.

In this work, we perform numerical simulations to study line force driven wind from a sub-Eddington luminosity thin disk in TDEs around a $10^6M_\odot$ black hole. We study how the wind properties change with decreasing accretion rate. The structure of this paper is as follows. In Section 2, we give detailed numerical settings of the simulations. In Section 3, we introduce the results. In Section 4, we discuss the results and give observational implication. We summarize the results in Section 5.

\section{Numerical method} \label{sec2.1}
After the disruption of the star, the fallback debris will form a accretion disc. The disc circularization and formation processes have been studied by many works (e.g., \citet{Hayasaki2013,Hayasaki2016,Shiokawa2015,Bonnerot2016,Steinberg2024,Price2024}). Generally, the disk circularization efficiency depends on many factors including the cooling efficiency, the black hole spin, the eccentricity of the debris orbit etc.  In this paper, we simply assume that after disruption, the fallback debris can very quickly be circularized and form accretion flow/disk around the black hole. If a star approaches the super massive black hole at galaxy center on a parabolic orbit, its mechanical energy (kinetic plus gravitational energy) is zero. We assume that the disrupted star moves on a parabolic orbit before disruption. We also assume that its orbit pericenter ($R_{\rm p}$) is equal to the tidal radius $R_{\rm T}$. We have an penetration factor $\beta = R_{\rm T}/R_{\rm p} = 1$. Under the condition that angular momentum is conserved, the outer boundary of the accretion disk ($R_{\rm C}$) is two times the pericenter.

In this paper, we assume that the black hole mass $M_{\rm BH} = 10^6 M_\odot$. The disrupted star is assumed to be a solar type star with mass $M_* = M_\odot$ and radius $R_* = R_\odot$. In this case, the tidal radius $R_{\rm T} = 47/2 R_{\rm s}$, with $R_{\rm s}$ being Schwarzschild radius. Base on the point of angular momentum conservation, we assume that after circularization of the fallback debris, an accretion disk forms inside $R_{\rm C} = 2R_{\rm T}/\beta =  47R_{\rm s}$ for $\beta = 1$. After disruption, when the debris fallback rate drops below the value of $\dot M_{\rm Edd} = L_{\rm Edd}/\eta c^2$ (with $L_{\rm Edd}$ and $\eta$ being Eddington luminosity and radiative efficiency of the accretion disk), it is believed that a thin disk forms around the black hole. In this paper, we assume that the radiative efficiency $\eta = 0.1$. We study the line force driven wind from the sub-Eddington accretion thin disk.

The fallback debris supplies gas to the accretion disk with a rate declining as $t^{-5/3}$, therefore, in reality, there is no steady state disk. When studying winds from a TDE accretion disk, one needs to trace a non-steady state accretion disk. However, we note that due to the fact that the cold thin accretion disk is geometrically very thin, it is hard to resolve the disk in numerical simulations. Therefore, in previous works studying line force driven wind from a thin disk, the thin disk is put at the mid-plane just as a boundary condition which supplies gas to form wind (e.g., \cite{Proga2000,Proga2004,Higginbottom2024}). We note that the thin disk evolution can be solved in a time-dependent one-dimensional self-similar approach. In this approach, it is found that the magnetic field and magnetically driven wind have strong effects on the time evolution of the accretion rate of the TDE thin disk \citep{Tamilan2024,Tamilan2025a,Tamilan2025b}.

Due to such technical difficulties, it is very hard to self-consistently simulate a thin disk in TDEs with a declining accretion rate. In this paper, we perform several discrete simulations. In each simulation, the thin disk is set to have a specific accretion rate. These simulations represent the accretion disk in TDEs at different accretion level. Very recently, \citet{Thomsen2022} use the same approach to study the evolution of the dynamics of TDE accretion flow in the super-Eddington accretion phase. In order to reduce computational resources, in \cite{Thomsen2022}, the authors also perform discrete simulations with different accretion rates. We note that in future, with developing of computational resources, it may becomes easier to self-consistently trace the time evolution of thin disk in TDEs.

\subsection{Basic equations} \label{sec2.2}
We perform two-dimensional axisymetric hydrodynamical simulations by using the PLUTO code \citep{Mignone2007}. We solve the below equations in spherical coordinates $(r, \theta , \phi)$,
\begin{equation}
\frac{d\rho}{dt} + \rho \nabla \cdot \bf{v} = 0
\end{equation}

\begin{equation}
\label{Ouler-equation}
\rho \frac{d {\bf v}}{dt} = -\nabla p -\rho \nabla \Phi + \rho {\bf F_{\rm rad}}
\end{equation}

\begin{equation}
\label{energy-equation}
\rho \frac{d}{dt} \left( \frac{e}{\rho} \right) = -p \nabla \cdot {\bf v} + \rho L_{\rm cool}
\end{equation}
Here, $\rho$, ${\bf v}$, $p$, $e$ are gas density, velocity, gas pressure, gas internal energy per unit volume, respectively. We employ an adiabatic equation of state $p = (\gamma-1) e$, with $\gamma= 5/3$. Pseudo-Newtonian $\Phi = -GM_{\rm BH}/(r-R_{\rm s})$ is used with $G$ being gravitational constant. We introduce the radiation pressure $\bf F_{\rm rad}$ and the net cooling rate $L_{\rm cool}$ below.

Our computational domain in radial direction is $10R_{\rm s} \leq r \leq 1500 R_{\rm s}$. In the $\theta$ direction, the domain covers $0^\circ \leq \theta \leq 90^\circ$. We note that previous simulations study line force driven wind usually set the inner computational boundary to be $30R_{\rm s}$ (e.g., \cite{Proga2000,Nomura2016}). Numerical simulations of both hot accretion flow (e.g., \cite{Yuan2012}) and super-Eddington accretion flow (e.g., \citet{Curd2019}) usually find that mass flux of wind is negligible compared to accretion rate on the black hole inside $10R_{\rm s}$. The reason for extremely weak wind inside $10R_{\rm s}$ is as follows. The gravity of black hole is so strong inside $10R_{\rm s}$ that the flow becomes supersonic moving to the horizon. Outside $10R_{\rm s}$, wind gradually becomes important with increasing radius. Therefore, the inner radial boundary in this paper is set to be $10R_{\rm s}$. The computational domain is dived into 174 grids in radial direction and 160 grids in $\theta$ direction. In order to well resolve the inner region, we use non-uniform grids in $r$ direction, with $\delta r_{i+1}/\delta r_{i} = 1.05$. In the $\theta$ direction, we use uniform grids.

In this paper, the method for calculating the line force driven wind is very similar as that of privious papers \citep{Proga2000,Nomura2017}. The main difference between this work and the previous works is that the thin disk at the mid-plane which emits mainly UV photons has an outer boundary of $47R_{\rm s}$. Below, we introduce the numerical settings in details.

At the sub-Eddington accretion phase of the accretion disk in TDEs, we assume that a thin disk is located at the mid-plane. Observations of AGNs show that a very compact hot corona radiating X-rays exists within 10 $R_{\rm s}$ \citep{Reis2013,Uttley2014}. We expect that at the sub-Eddington accretion phase in TDEs, there may exist a similar X-ray radiating corona. Therefore, in our simulations, we also assume that a hot corona exists inside 10 $R_{\rm s}$, which radiates isotropic X-ray photons. The luminosity of the thin disk is described as $L_{\rm D} = \varepsilon L_{\rm Edd}$, where $\varepsilon$ is the Eddington ratio of the disk luminosity. The luminosity of the X-ray corona source is $L_{\rm X} = f_{\rm X}L_{\rm D}$. \cite{Li2019} studied a sample of luminous AGNs. Generally, they found that the ratio of the X-ray luminosity to the bolometric luminosity decreases with the Eddington ratio $\varepsilon$. In their table 2, for the AGNs with Eddington ratios in the range 0.3-1.0, the value of $f_{\rm X}$ is in the range of 0.015-0.05. Therefore, in our paper, we choose a fiducial value of $f_{\rm X} = 0.03$. We do tests of varying the value of $f_X$ in Appendix C. The radiation intensity of the thin disk is
\begin{equation}
    I_{\rm D} (r_{\rm D}) = \frac{\varepsilon L_{\rm Edd}}{12 \pi^2 R_{\rm s}^2} \frac{27R_{\rm s}^3}{r_{\rm D}^3} \left[ 1-\left( \frac{3R_{\rm s}}{r_{\rm D}}\right)^{1/2} \right]
\end{equation}
where $r_{\rm D}$ is the radius of the thin disk. We note that the inner boundary of the thin disk is $3R_{\rm s}$.
The effective temperature of the thin disk is,
\begin{equation}
T_{\rm eff} (r_{\rm D}) =  \left( \pi \frac{I_{\rm D} (r_{\rm D})}{\sigma} \right)^{1/4}
\end{equation}
where $\sigma$ is the Stefan-Boltzmann parameter.

At a specific location in the computational domain ($r$, $\theta$), if we neglect the optical depth effect, the radiation flux from the thin disk at mid-plane is calculated
\begin{equation}
    {\bf F}_{\rm D, thin}(r, \theta) = \oint I({\bf{r},{\hat{n}})} {\hat{n}} d \Omega
\end{equation}
where $\hat{n}$ is the unit vector, and $d \Omega$ is the solid angle subtended by the disk at the midplane, $\bf{r}$ is the position vector. We refer to the Appendix of \cite{Proga1998} for a detailed calculation of radiation from the thin disk. The attenuation of the disk emission is set to be due to electron scattering which is $\kappa_{\rm es} = 0.34 {\rm g^{-1} cm^2}$. Therefore, the radiation flux from the thin disk is $F_{\rm D} = F_{\rm D, thin}e^{-\tau_{\rm UV}}$, with $\tau_{\rm UV}=\int\kappa_{\rm es}\rho d{\bf r}$. The attenuation of X-ray photons from the corona depends on the ionization parameter $\xi$, which is defined as $\xi = L_{\rm X}/n r^2$, with $n$ being gas number density. Following \cite{Proga2000}, if $\xi \geq 10^5 {\rm erg\ s^{-1} \ cm}$, the X-ray attenuation is set to be equal to $\kappa_{\rm X} = \kappa_{\rm es}$. If  $\xi < 10^5 {\rm erg\ s^{-1} \ cm}$, the X-ray attenuation is set to be $\kappa_{\rm X} = 100 \kappa_{\rm es}$.

In Equation (\ref{Ouler-equation}), the radiation pressure on unit pass is
\begin{equation}
    {\bf F_{\rm rad}} = \frac{\sigma_{\rm T}{\bf F_{\rm D}}}{m_{\rm p} c} + \int_\Omega \mathcal{M}\frac{\sigma_{\rm T} I({\bf{r}},{\hat{n}})}{m_{\rm p}c} {\hat{n}} d \Omega
\end{equation}
where $\sigma_T$ is the Thomson scattering cross section, $m_{\rm p}$ is the proton mass. The first term corresponds to radiation force due to electron scattering. The second term is the line force. In the second term, $\mathcal{M}$ is the force multiplier, which is defined as the ratio of the line force to the radiation force due to electron scattering. The force multiplier $\mathcal{M}$ is a function of the ionization parameter $\xi$ and the local optical depth parameter \citep{Rybicki1978}. We introduce the details of the force multiplier in Appendix A. We also refer to Equations (11) -(16) in \cite{Proga2000} for the details of force multiplier $\mathcal{M}$. Because, the X-ray flux is significantly smaller than the radiation flux from the thin disk, we neglect its contribution to the radiation pressure. We have done tests and found that including the radiation pressure due to the X-rays will not affect the results.

The radiative cooling rate $L_{\rm cool}$ in Equation (\ref{energy-equation}) includes Compton heating/cooling, photoionization heating-recombination cooling, bremsstrahlung cooling and line cooling. Both the Compton heating/cooling and photoionization heating-recombination cooling is due to the interaction of the X-ray from the hot corona and gas. As down in \cite{Proga2000}, for the X-rays, we set a 10 keV bremsstrahlung spectrum, which has a Compton temperature $T_{\rm X} = 10^8 {\rm K}$. We refer to Equations (18)-(21) in \cite{Proga2000} for the details of cooling function $L_{\rm cool}$.

\subsection{Initial and boundary conditions} \label{sec2.3}

\begin{figure}[ht]
\includegraphics[width=0.48\textwidth]{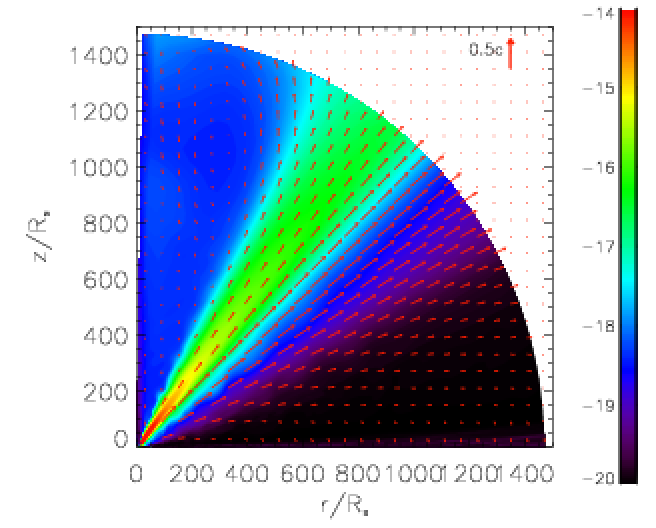}
       \ \centering \caption{Time-averaged logarithm gas density (color) in ${\rm g \ cm^{-3}}$ and velocity (vectors) for $\varepsilon=0.8$. The $z$-axis is the rotational axis of the accretion disk. The accretion disk surface is at the $z=0$ plane.}\label{fig:vector}
\end{figure}

\begin{figure}[ht]
\includegraphics[width=0.48\textwidth]{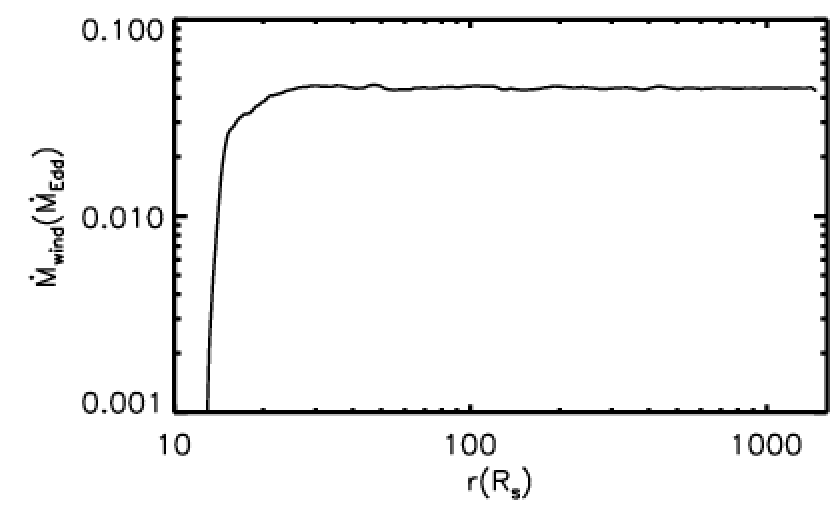}
\includegraphics[width=0.48\textwidth]{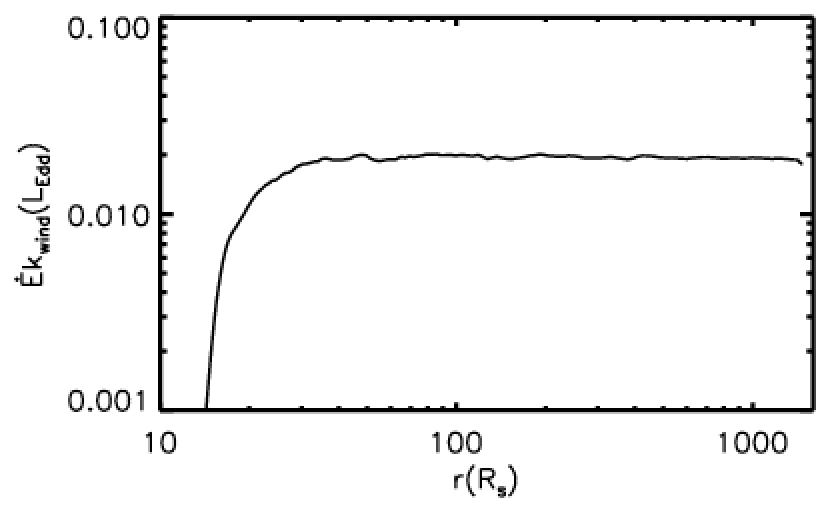}
\includegraphics[width=0.48\textwidth]{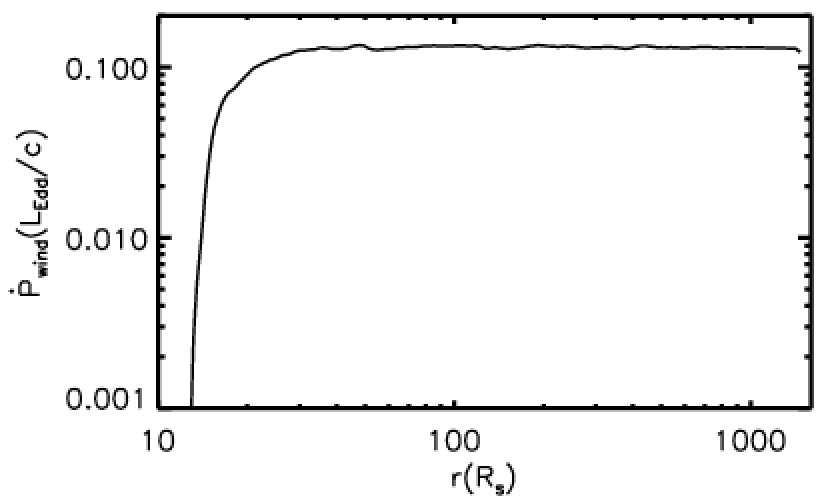}
       \ \centering \caption{Time-averaged radial profile of wind mass flux (top-panel), kinetic power (middle panel) and momentum flux (bottom panel) for the model with $\varepsilon = 0.8$.}\label{fig:radialprofile}
\end{figure}

The location $\theta=90^\circ$ corresponds to the location of the surface of the standard thin accretion disk. Following \cite{Nomura2017}, we set the surface of the thin disk as follows   \citep{Shakura1973,Kato2008,Nomura2017} \footnote{We give the derivation of the scaling law for thin disk density in Equation 8 in Appendix B.},
\begin{equation*}
\rho(\theta=\pi/2) =
5\times10^{-4}(M_{\rm BH}/M_\odot)^{-1} (\varepsilon/\eta)^{-2}(r/R_{\rm s})^{3/2} {\rm g \ cm^{-3}}
\end{equation*}
\begin{equation}
\ \ \ \ \ \ \ \ \ \ \ \ \ \ \ \ \ \ \ \ \ \ \ \ \ \ \ \ \ \ \ \ \ \ \ \ \ \ \ \ \ \ \ \ \ \ \ \ \ \ \ \ \ \ \ \ \ \ r \leq 47R_{\rm s}
\end{equation}
\begin{equation}
\rho(\theta=\pi/2) = \rho_0 \ \ \ \ \ \ \ \ \ \ \ \ \ \ \ \ \ \ \ \ \ \ \ \ \ \ \ \ \  \ \ \ \ \ \ \ \ \ \ \ \ \ \ \ \ \ \ \ \ \  \ \ \ r > 47R_{\rm s}
\end{equation}
The outer boundary for the accretion disk around a $10^6M_\odot$ in TDEs is $47R_{\rm s}$. Therefore, outside $47R_{\rm s}$, we set a quite low density $\rho_0$, which equals to $10^{-11} \rho (r=30R_{\rm s}, \theta = \pi/2)$. At $\theta=\pi/2$, for the disk surface inside $47R_{\rm s}$, the radial velocity and the rotational velocity are always set to be 0 and the Keplerian value, respectively. For $v_\theta$, we initially set it to be 0.

Above the thin disk inside $47R_{\rm s}$, we assume hydrostatic equilibrium in the vertical direction, then the initial density distribution is
\begin{equation}
    \rho(r,\theta) = \rho(r,\pi/2)\exp \left( -\frac{GM_{\rm BH}}{2c_{\rm s}^2r\tan^2(\theta)} \right)
\end{equation}
where $c_{\rm s}$ is the sound speed at the disk surface. In order to avoid numerical difficulty, we set a density floor, which is equal to $\rho_0$. The value of $\rho_0$ is negligibly small compared to the density inside $47R_{\rm s}$. Therefore, the density floor can not affect the properties of wind. The initial temperature at a location $(r, \theta)$ is set to be $T(r,\theta) = T_{\rm eff}(r\sin\theta)$.

In the region $\theta < \pi/2$, the initial velocity $v_r = v_\theta =0$; the rotational velocity is set to balance the black hole gravity.

At the radial inner and outer boundary, we use outflow boundary conditions. At the accretion disk rotational axis $\theta=0$, we use axially symmetric boundary conditions.

\begin{figure}[ht]
\includegraphics[width=0.48\textwidth]{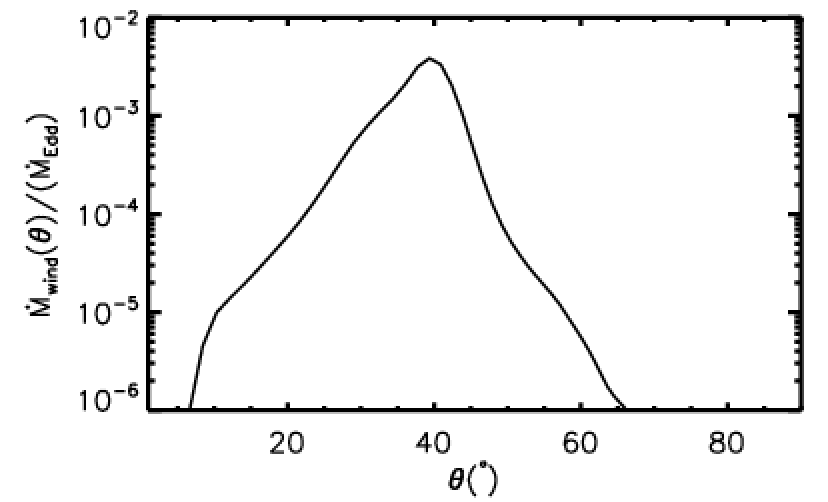}
\includegraphics[width=0.48\textwidth]{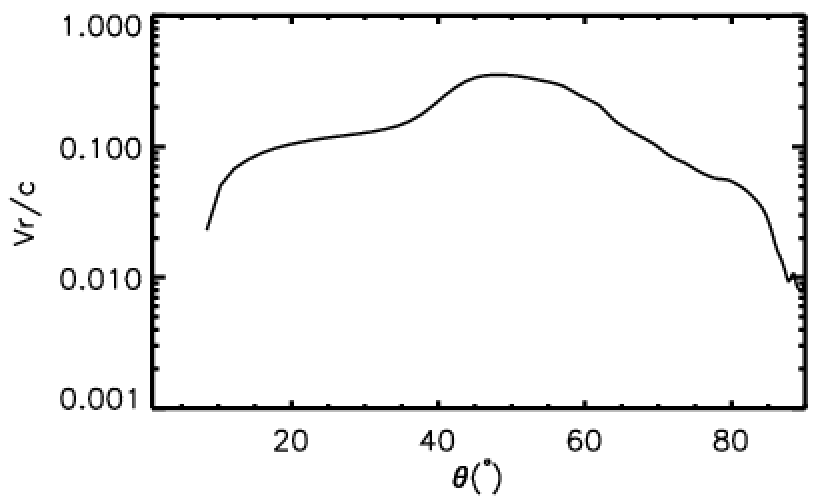}
       \ \centering \caption{Time-averaged angular profile of wind mass flux (top-panel), radial velocity (bottom panel) measured at the outer radial boundary for the model with $\varepsilon = 0.8$.}\label{fig:angularprofile}
\end{figure}

\section{Result} \label{sec3}
\subsection{Overview of structure of line force driven wind}
We take the simulation with disk luminosity of Eddington ratio $\varepsilon = 0.8$ as our fiducial model. Figure \ref{fig:vector} shows the time-averaged gas density (color) and poloidal velocity (arrows). From this figure, it is clear that although the accretion disk is just present inside $47R_{\rm s}$, strong wind can be launched from the small disk. The main stream with high density and velocity occurs in the angular range $28^\circ < \theta < 45^\circ$. Below, we introduce the properties of wind in details.

We calculate the mass flux, kinetic power and momentum flux of wind as a function of radius as follows,
\begin{equation}
    \dot M_{\rm wind} (r) = 4\pi r^2\int^{{90}^\circ}_0 \rho v_r \sin \theta {\rm d} \theta
\end{equation}

\begin{equation}
    \dot E_{\rm K wind} (r) = 4\pi r^2\int^{{90}^\circ}_0 \frac{1}{2}\rho v_r^3 \sin \theta {\rm d} \theta
\end{equation}

\begin{equation}
    \dot P_{\rm wind}(r) = 4\pi r^2\int^{{90}^\circ}_0 \rho v_r^2 \sin \theta {\rm d} \theta
\end{equation}
The results are shown in Figure \ref{fig:radialprofile}. The wind is mainly launched and accelerated inside $30R_{\rm s}$. Outside $30R_{\rm s}$, all the fluxes are constant with radius. The results demonstrate that the wind launching and acceleration processes are finished inside $30R_{\rm s}$. \cite{Nomura2017} performed simulations to study line force driven wind from an accretion disk in AGNs. They found that the wind in their simulations is launched mainly in the region inside $40R_{\rm s}$, which is quite similar as the case in the present paper. We do test of the impact of the TDE disk outer radius in Appendix C. We find that changing the disk outer radius would not affect the results much. The reason may be as follows. The wind is produced by two steps \citep{Nomura2017}. First, the gas needs to be puffed up from the disk by radiation pressure exerted by local UV photons. Second, the puffed up gas is radially accelerated by line force to form wind. We find that for a standard thin disk around a $10^6 M_\odot$ black hole, considering Eddington ratio around $\varepsilon=0.5$, the UV photons from the region within $40R_{\rm s}$ contribute to $80\%$ of the total disk UV emission. Therefore, the disk gas inside $40R_{\rm s}$ can be puffed up by radiation pressure much easier. Outside $40R_{\rm s}$, the UV disk emission is weak and disk gas is much hard to be puffed up to form wind. Therefore, despite that the disk size in AGN is significantly larger than that of the disk in TDEs studied in the present work, they \citep{Nomura2017} found that the line force driven wind is mainly launched very near the black hole too.

\begin{figure}[ht]
\includegraphics[width=0.48\textwidth]{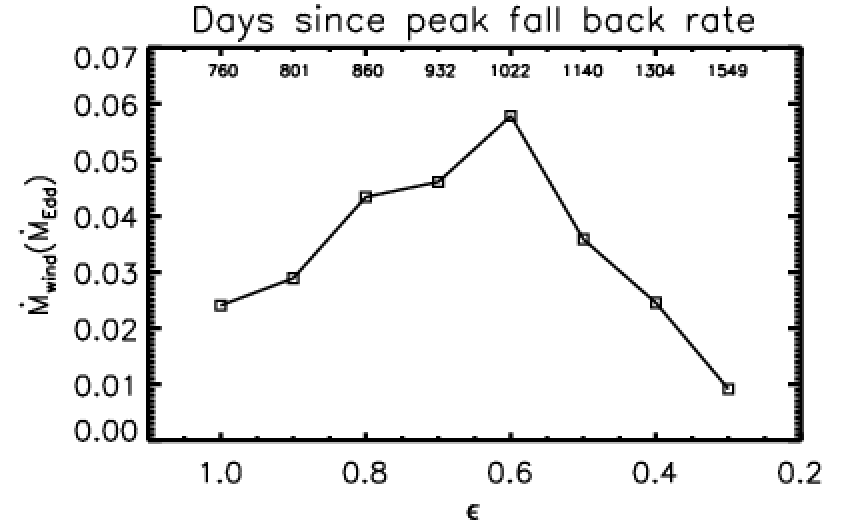}
\includegraphics[width=0.48\textwidth]{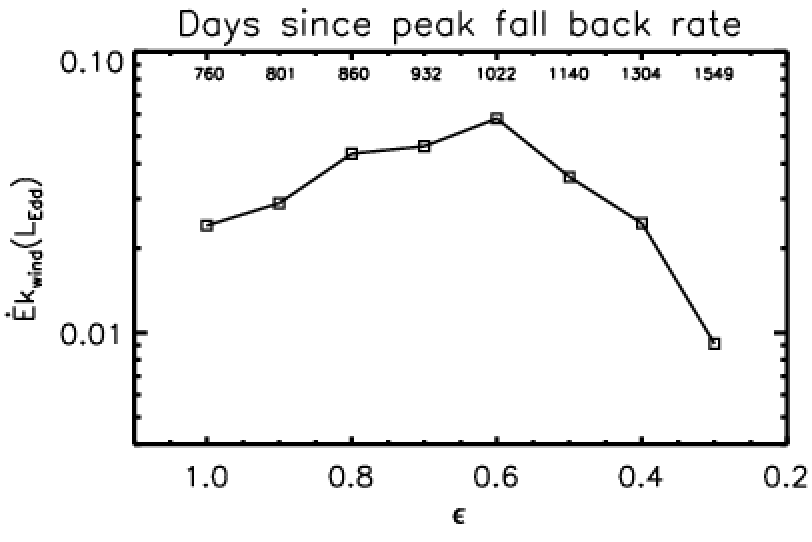}
       \ \centering \caption{Mass flux (top-panel) and kinematic power (bottom panel) of wind measured at the outer radial boundary as a function of Eddington ratio.}\label{fig:timeevolution}
\end{figure}

The mass flux of wind is $4.34\%\dot M_{\rm Edd}$. Because the mass accretion rate is $80\%\dot M_{\rm Edd}$, the ratio of the mass flux of wind to the mass accretion rate is $5.4\%$. In the sub-Eddington accretion phase, more than $94\%$ of the fallback stellar debris is accreted by the black hole. For comparison, in the super-Eddington accretion flow around a $10^6M_\odot$ black hole in TDEs, $43\%$ of the fallback debris is accreted \citep{Bu2023b}.

The kinematic power of wind is $\sim 2.3 \times 10^{42} {\rm erg/s}$. For comparison, in the super-Eddington phase, the kinematic power of wind is well above $10^{44} {\rm erg/s}$ \citep{Bu2023b}. The momentum flux of wind is $12\%$ of that of radiation of the thin disk, which is consistent with the fact that for radiation pressure driven wind, its momentum flux is smaller than that of radiation.

We show the angular profile of mass flux and radial velocity of wind at the radial outer boundary in Figure \ref{fig:angularprofile}. It is clear that the main stream of wind is in the angular range of $28^\circ < \theta < 45^\circ$ as shown in Figure \ref{fig:vector}. The maximum radial velocity can be as high as $0.3c$ and is present around $\sim 45^\circ $. The minimum wind velocity is $\sim 0.01c$.

\subsection{Eddington ratio dependence}

We have also run simulations with different values of $\varepsilon$. We find that when $\varepsilon \leq 0.2$, the line force is too weak to launch wind. \cite{Proga2000} and \cite{Proga2004} also found that when the Eddington ratio of the accretion disk is as low as $0.1$, winds do not appear.

Generally, the angular distributions of mass flux, the angular profile of wind velocity in models with different values of $\varepsilon$ are quite similar as those shown in Figure \ref{fig:angularprofile}. We show the mass flux and kinematic power of wind in TDEs as a function of time (or mass accretion rate) in Figure \ref{fig:timeevolution}. The top horizontal axis in this figure shows the time after peak debris fallback rate. The time is calculated according to the equation $\dot M_{\rm fallback} = 1/3(M_*/t_{\rm fb})(t/t_{\rm fb})^{-5/3}$. For the parameters $M_{\rm BH} = 10^6M_\odot$ and $M_*=M_\odot$ used in this work, the transition from the super-Eddington fallback phase to sub-Eddington fallback phase occurs at $t_{\rm cr} =760$ days.

From 760 to 1549 days after peak fallback rate, the mass flux of wind is in the range of $1\%-6\% \dot M_{\rm Edd}$. The kinematic power of wind is in the range of $1\%-6\% L_{\rm Edd}$. The wind achieves its most strong value when $\varepsilon=0.6$. The reason is as follows. With the increase of $\varepsilon$, the X-ray photon flux which can photo-ionize gas increases, which will result in high ionization parameter of gas. The line force multiplier $\mathcal{M}$ decreases with increasing ionization parameter. Therefore, when $\varepsilon \geq 0.6$, strength of wind decreases with increase of $\varepsilon$. When $\varepsilon < 0.6$, the strength of wind decreases with decreasing $\varepsilon$. The reason is that smaller $\varepsilon$ results in a smaller radiation flux and radiation pressure.

We time-integrate the mass and kinetic energy taken away by wind from $t=760$ to $1549$ days. We find that during this period, the mass and kinetic energy taken away by wind are $0.15\% M_\odot$ and $1.6 \times 10^{50} {\rm erg}$, respectively.

\section{Discussion and observational applications}

The magnetic driven wind mechanism is neglected in the present paper. The magnetic field should affect the evolution of the TDE disk significantly. For example, very recently \cite{Tamilan2025b} find that a strong magnetic field can make the accretion rate evolves as a $t^{-5/2}$ law. Therefore, in future, a $t^{-5/2}$ law decline of radiation in observations of TDEs may help to identifying the TDEs disk being magnetic field dominated. In order to study magnetically driven wind, the geometry and strength of the magnetic field must be known. When magnetic field effects are included, the properties of the line force driven wind may change. \cite{Yang2021a} found that when a weak poloidal large-scale magnetic field is present, the wind velocity and covering factor will both become larger. The reason is that in the presence of magnetic field, the region around the rotational axis becomes magnetic pressure dominated, which prevents gases from spreading to higher latitudes and then enhances the gas column
density at middle and low latitudes. Higher column density is helpful to shield X-ray photons, which
causes the line force to be more effective in driving wind. It is not clear how the properties of line force driven wind will be in the presence of small-scale tangled magnetic field. In reality, both magnetic field and line force play roles in driving wind. Therefore, in future, it is important to study the line force driven wind in the presence of magnetic field in the context of TDEs.

The interaction of wind in TDEs and the CNM is believed to be the origin of radio emission in some TDEs \citep{Alexander2020}. Here, we estimate the possible radio emission from the interaction of the line force driven wind and the CNM. The wind-CNM interaction can accelerate non-thermal electrons. We follow the model of \cite{Matsumoto2021} to calculate the radio emission luminosity. We take the wind primarily launched within the main stream of polar angle range of $\approx28^{\circ}-45^{\circ}$, with an average velocity of $0.2$ c. We calculate the radio emission $1000$ days after the wind launching. For the CNM, we take the model-dependent density profile of AT2019dsg given by \cite{Matsumoto2024}. The fractions of shocked thermal energy transferred to non-thermal electrons and magnetic fields are set to $\epsilon_e = \epsilon_B = 0.1$ with a non-thermal electron spectral index $p=2.5$. In this scenario, the synchrotron self-absorption frequency is calculated as $\nu_{a} \sim 3.4$ GHz and the corresponding synchrotron luminosity is $\nu L_{\nu}\sim 10^{40}$ erg s$^{-1}$. The radio luminosity calculated here is sufficient to account for the observed radio emissions from TDEs \citep{Alexander2020}. Therefore, the line force driven wind is capable to generate radio emission via wind-CNM interaction, which can be observed by current telescopes. In future, it is very necessary to study in details about the related radio emission.

Cold cloud may exist around the SMBHs at galaxy centers. The observations to our Galaxy center show a cold and very dense circumnuclear disk at the parsec scale \citep{Zhao2016} and an inner `mini-spiral' \citep{Tsuboi2016}. Therefore, it seems that cold cloud and hot diffuse CNM can coexist near quiescent SMBHs. Wind can collides with the cold dense cloud surrounding the SMBH. The bow shocks can be induced, which will accelerate power-law electrons \citep{Mou2022,Bu2023a}. In the wind-cloud model, the radio light curve can have very steep rise \citep{Mou2022}. Also, if the cloud distribution is quite non-uniform, one can predict strong fluctuations of the radio emission. These features are different from those of the wind-CMN model.
We estimate the radio emission in this scenario. We adopt the same parameters as in the above outflow-CNM colliding model, including: fractions of shocked thermal energy transferred to non-thermal electrons and magnetic fields ($\epsilon_e = \epsilon_B = 0.1$). We assume wind velocity to be $0.2c$. We calculate the radio emission 1000 days after the wind launching. The cloud location is $r \approx v_{\rm wind}T = 5.2 \times 10^{17}$ cm, with cloud size ($R_{\rm cloud} =\eta_{\rm cloud} r$). The covering factor of cloud is assumed to be $c_{f} = 0.2$. The wind is primarily launched  within a polar angle range of $\approx28^{\circ}$-$45^{\circ}$, with a kinetic power $L_{\rm kin} \approx 2.3 \times 10^{42}$ erg s$^{-1}$. Under these conditions, we find that when $\eta_{\rm cloud} = 0.001$, the synchrotron self-absorption frequency is $\nu_{a} = 0.1$ GHz, and the corresponding radio luminosity $\nu L_{\nu} = 4 \times 10^{35}$ erg s$^{-1}$. For $\eta_{\rm cloud} = 0.1$, we have $\nu_a = 0.5$ GHz, and the radio luminosity is $\nu L_\nu = 6\times 10^{37}$ erg s$^{-1}$. For cold cloud with much larger size ($\eta_{\rm cloud} = 0.1$), the radio emission via wind-cloud interaction can be as high as those observed by current telescopes \citep{Alexander2020}.

\section{Summary} 

We study the line force driven wind from the sub-Eddington accretion disk around a $10^6M_\odot$ black hole in TDEs. In our models, the accretion disk has a size of two times the pericenter radius, which is $47R_{\rm s}$ by assuming the penetration factor $\beta = 1$ and a solar type star being disrupted.

We find that although the disk size is small, strong wind can be driven when the disk luminosity has Eddington ratio $\varepsilon > 0.2$. The wind is mainly launched and accelerated inside $30R_{\rm s}$. Outside $30R_{\rm s}$. The mass flux, kinetic power and momentum flux of wind are all a constant with radius. The maximum velocity of wind can be as high as $0.3c$. The disk wind has mass flux in the range of $ 1\%-6\% \dot M_{\rm Edd}.$  The kinematic power of wind is in the range of $1\%-6\% L_{\rm Edd}$. The possible radio emission from the wind-CNM/clouds interaction is briefly discussed.

\begin{acknowledgements}
D. Bu is supported by the Natural Science Foundation of China (grants 12173065, 12133008, 12192220, 12192223). X. Yang is supported by Chongqing Natural Scince Foundation (grant CSTB2023NSCQ-MSX0093) and the Natural Science Foundation of China (grant 12347101). L. Chen is supported by NSFC (12173066), National Key R\&D program of China (2024YFA1611403), National SKA Program of China (2022SKA0120102) and Shanghai Pilot Programme for Basic Research, CAS Shanghai Branch (JCYJ-SHFY-2021-013). G. Mou is supported by the NSFC (grants 12473013, 12133007).
\end{acknowledgements}

\bibliographystyle{aa}
\bibliography{reference}

\begin{thebibliography}{100}
\expandafter\ifx\csname natexlab\endcsname\relax\def\natexlab#1{#1}\fi

\bibitem[{{Alexander} {et~al.}(2016){Alexander}, {Berger}, {Guillochon},
  {Zauderer}, \& {Williams}}]{Alexander2016}
{Alexander}, K.~D., {Berger}, E., {Guillochon}, B.~A., {Zauderer}, B.~A., \&
  {Williams}, P. K.~G. 2016, ApJ, 819, L25

\bibitem[{{Alexander} {et~al.}(2020){Alexander}, {van Velzen}, \&
  {Horesh}}]{Alexander2020}
{Alexander}, K.~D., {van Velzen}, S., \& {Horesh}, A.~{Zauderer}, B.~A. 2020,
  SSRv, 216, 81

\bibitem[{{Barniol Duran} {et~al.}(2013){Barniol Duran}, {Nakar}, \&
  {Prian}}]{Barniol2013}
{Barniol Duran}, R., {Nakar}, E., \& {Prian}, T. 2013, ApJ, 772, 78

\bibitem[{{Blandford} \& {Payne}(1982)}]{Blandford1982}
{Blandford}, R.~D. \& {Payne}, D.~G. 1982, MNRAS, 199, 883

\bibitem[{{Bonnerot} {et~al.}(2016){Bonnerot}, {Rossi}, {Lodaot}, \&
  {Price}}]{Bonnerot2016}
{Bonnerot}, C., {Rossi}, E.~M., {Lodaot}, G., \& {Price}, D.~J. 2016, MNRAS,
  455, 2253

\bibitem[{{Bu} {et~al.}(2023{\natexlab{a}}){Bu}, {Chen}, {Mou}, {Qiao}, \&
  {Yang}}]{Bu2023a}
{Bu}, D., {Chen}, L., {Mou}, G., {Qiao}, E., \& {Yang}, X. 2023{\natexlab{a}},
  MNRAS, 521, 4180

\bibitem[{{Bu} {et~al.}(2023{\natexlab{b}}){Bu}, {Qiao}, \& {Yang}}]{Bu2023b}
{Bu}, D., {Qiao}, E., \& {Yang}, X. 2023{\natexlab{b}}, MNRAS, 523, 4136

\bibitem[{{Bu} {et~al.}(2022){Bu}, {Qiao}, {Yang}, {Liu}, {Chen}, \&
  {Wu}}]{Bu2022}
{Bu}, D., {Qiao}, E., {Yang}, X., {et~al.} 2022, MNRAS, 516, 2833

\bibitem[{{Cannizzo} {et~al.}(1990){Cannizzo}, {Lee}, \&
  {Goodman}}]{Cannizzo1990}
{Cannizzo}, J.~K., {Lee}, H.~M., \& {Goodman}, J. 1990, ApJ, 351, 38

\bibitem[{{Castor} {et~al.}(1975){Castor}, {Abbott}, \& {Klein}}]{Castor1975}
{Castor}, J.~I., {Abbott}, D.~C., \& {Klein}, R.~I. 1975, ApJ, 195, 157

\bibitem[{{Cendes} {et~al.}(2021){Cendes}, {Alexander}, {Berger}, {Eftekhari},
  {Williams}, \& {Chornock}}]{Cendes2021}
{Cendes}, Y., {Alexander}, K.~D., {Berger}, E., {et~al.} 2021, ApJ, 919, 127

\bibitem[{{Cendes} {et~al.}(2024){Cendes}, {Berger}, {Alexander}, {Chornock},
  {Margutti}, {Metzger}, \& {Wieringa}}]{Cendes2024}
{Cendes}, Y., {Berger}, E., {Alexander}, K.~D., {et~al.} 2024, ApJ, 971, 185

\bibitem[{{Cendes} {et~al.}(2022){Cendes}, {Berger}, {Alexander}, {Hajela}, \&
  {Chornock}}]{Cendes2022}
{Cendes}, Y., {Berger}, E., {Alexander}, K. D.~{Gomez}, S., {Hajela}, A., \&
  {Chornock}, R. 2022, ApJ, 938, 28

\bibitem[{{Cufari} {et~al.}(2022){Cufari}, {Coughlin}, \& {Nixon}}]{Cufari2022}
{Cufari}, M., {Coughlin}, E.~R., \& {Nixon}, C.~J. 2022, ApJ, 924, 34

\bibitem[{{Curd} \& {Narayan}(2019)}]{Curd2019}
{Curd}, B. \& {Narayan}, R. 2019, MNRAS, 483, 565

\bibitem[{{Curd} \& {Narayan}(2023)}]{Curd2023}
{Curd}, B. \& {Narayan}, R. 2023, MNRAS, 518, 3441

\bibitem[{{Dai} {et~al.}(2018){Dai}, {McKinney}, {Roth}, {Ramirez-Ruiz}, \&
  {Miller}}]{Dai2018}
{Dai}, L., {McKinney}, J.~C., {Roth}, N., {Ramirez-Ruiz}, E., \& {Miller},
  M.~C. 2018, ApJ, 859, L20

\bibitem[{{Evans} \& {Kochanek}(1989)}]{Evans1989}
{Evans}, C.~R. \& {Kochanek}, C.~S. 1989, ApJ, 346, L13

\bibitem[{{Frank} {et~al.}(1992){Frank}, {King}, \& {Raine}}]{Frank1992}
{Frank}, J., {King}, A., \& {Raine}, D. 1992, Camb. Astrophys. Ser., Vol. 21,
  Accretion power in astrophysics

\bibitem[{{Fukumura} {et~al.}(2015){Fukumura}, {Tombesi}, {Kazanas}, {Shrader},
  {Behar}, \& {Contopoulos}}]{Fukumura2015}
{Fukumura}, K., {Tombesi}, F., {Kazanas}, D., {et~al.} 2015, ApJ, 805, 17

\bibitem[{{Gezari}(2021)}]{Gezari2021}
{Gezari}, S. 2021, ARA\&A, 59, 21

\bibitem[{{Golightly} {et~al.}(2019){Golightly}, {Coughlin}, \&
  {Nixon}}]{Golightly2019}
{Golightly}, E. C.~A., {Coughlin}, E.~R., \& {Nixon}, C.~J. 2019, ApJ, 872, 163

\bibitem[{{Goodwin} {et~al.}(2022){Goodwin}, {van Velzen}, {Miller-Jones},
  {Bietenholz}, {ZWederfoort}, {Hammerstein}, {Bonnerot}, {Hoffmann}, \&
  {Yan}}]{Goodwin2022}
{Goodwin}, A.~J., {van Velzen}, S., {Miller-Jones}, J. C. A.~{Mummery}, A.,
  {et~al.} 2022, MNRAS, 511, 5328

\bibitem[{{Guillochon} {et~al.}(2016){Guillochon}, {McCourt}, {Chen},
  {Johnson}, \& {Berger}}]{Guillochon2016}
{Guillochon}, J., {McCourt}, M., {Chen}, X., {Johnson}, M.~D., \& {Berger}, E.
  K.~G. 2016, ApJ, 822, 48

\bibitem[{{Guo} {et~al.}(2025){Guo}, {Sun}, {Li}, {Jiang}, {Wang}, {Bu},
  {Jiang}, {Wang}, {Yao}, {Shen}, {Gu}, \& {Sun}}]{Guo2025}
{Guo}, H., {Sun}, J., {Li}, S., {et~al.} 2025, ApJ, 979, 235

\bibitem[{{Hayasaki} {et~al.}(2013){Hayasaki}, {Stone}, \&
  {Loeb}}]{Hayasaki2013}
{Hayasaki}, K., {Stone}, N., \& {Loeb}, A. 2013, MNRAS, 434, 909

\bibitem[{{Hayasaki} {et~al.}(2016){Hayasaki}, {Stone}, \&
  {Loeb}}]{Hayasaki2016}
{Hayasaki}, K., {Stone}, N., \& {Loeb}, A. 2016, MNRAS, 461, 3760

\bibitem[{{Hayasaki} \& {Yamazaki}(2023)}]{Hayasaki2023}
{Hayasaki}, K. \& {Yamazaki}, R. 2023, ApJ, 954, 5

\bibitem[{{Hayasaki} {et~al.}(2018){Hayasaki}, {Zhong}, {Li}, {Berczik}, \&
  {Spurzem}}]{Hayasaki2018}
{Hayasaki}, K., {Zhong}, S., {Li}, S., {Berczik}, P., \& {Spurzem}, R. 2018,
  ApJ, 855, 129

\bibitem[{{Higginbottom} {et~al.}(2024){Higginbottom}, {Scepi}, {Nicolas},
  {Long}, {Matthews}, \& {Sim}}]{Higginbottom2024}
{Higginbottom}, N., {Scepi}, N., {Nicolas}, K.~C., {et~al.} 2024, MNRAS, 527,
  9236

\bibitem[{{Horesh} {et~al.}(2021){Horesh}, {Cenko}, \& {Arcavi}}]{Horesh2021}
{Horesh}, A., {Cenko}, S.~B., \& {Arcavi}, I. 2021, NatAs, 5, 491

\bibitem[{{Huang} {et~al.}(2024){Huang}, {Davis}, \& {Jiang}}]{Huang2024}
{Huang}, X., {Davis}, S.~W., \& {Jiang}, Y. 2024, ApJ, 974, 165

\bibitem[{{Hung}(2017)}]{Hung2017}
{Hung}, T.~e.~a. 2017, ApJ, 842, 29

\bibitem[{{Jiang} {et~al.}(2016){Jiang}, {Guillochon}, \& {Loeb}}]{Jiang2016}
{Jiang}, Y., {Guillochon}, J., \& {Loeb}, A. 2016, ApJ, 830, 125

\bibitem[{{Kato} {et~al.}(1978){Kato}, {Fukue}, \& {Mineshige}}]{Kato2008}
{Kato}, S., {Fukue}, J., \& {Mineshige}, S. 1978, Black-Hole Accretion Disks.
  Kyoto University Press, Kyoto, Japan

\bibitem[{{Komossa}(2015)}]{Komossa2015}
{Komossa}, S. 2015, JHEAp, 7, 148

\bibitem[{{Krolik} {et~al.}(2016){Krolik}, {Piran}, {Svirski}, \&
  {Cheng}}]{Krolik2016}
{Krolik}, J., {Piran}, T., {Svirski}, G., \& {Cheng}, R.~M. 2016, ApJ, 827, 127

\bibitem[{{Lei} {et~al.}(2024){Lei}, {Wu}, {Li}, {Li}, {Lei}, {Fan}, {Wu},
  {Wang}, \& {Yang}}]{Lei2024}
{Lei}, X., {Wu}, Q., {Li}, H., {et~al.} 2024, ApJ, 977, 63

\bibitem[{{Li} \& {Cao}(2022)}]{Li2022}
{Li}, J. \& {Cao}, X. 2022, ApJ, 926, 11

\bibitem[{{Li}(2019)}]{Li2019}
{Li}, S. 2019, MNRAS, 490, 3793

\bibitem[{{Li} \& {Begelman}(2014)}]{Li2014}
{Li}, S. \& {Begelman}, M.~C. 2014, ApJ, 786, 6

\bibitem[{{Liu} {et~al.}(2021){Liu}, {Cao}, {Abramowicz}, {Cao}, \&
  {Zhou}}]{Liu2021}
{Liu}, F., {Cao}, C., {Abramowicz}, M. A.~{Wielgus}, M., {Cao}, R., \& {Zhou},
  Z. 2021, ApJ, 908, 179

\bibitem[{{Liu} {et~al.}(2017){Liu}, {Zhou}, {Cao}, {Ho}, \&
  {Komossa}}]{Liu2017}
{Liu}, F., {Zhou}, Z., {Cao}, R., {Ho}, L.~C., \& {Komossa}, S. 2017, MNRAS,
  472, L99

\bibitem[{{Lodato} {et~al.}(2009){Lodato}, {King}, \& {Pringle}}]{Lodato2009}
{Lodato}, G., {King}, A.~R., \& {Pringle}, J.~E. 2009, MNRAS, 392, 332

\bibitem[{{Lodato} \& {Rossi}(2011)}]{Lodato2011}
{Lodato}, G. \& {Rossi}, E.~M. 2011, MNRAS, 410, 359

\bibitem[{{Lu} \& {Bonnerot}(2020)}]{Lu2020}
{Lu}, W. \& {Bonnerot}, C. 2020, MNRAS, 492, 686

\bibitem[{{Lynden-Bell}(1996)}]{Lynden-Bell1996}
{Lynden-Bell}, D. 1996, MNRAS, 279, 389

\bibitem[{{Mageshwaran} {et~al.}(2023){Mageshwaran}, {Shaw}, \&
  {Bhattacharyya}}]{Mageshwaran2023}
{Mageshwaran}, T., {Shaw}, G., \& {Bhattacharyya}, S. 2023, MNRAS, 518, 5693

\bibitem[{{Matsumoto} \& {Piran}(2021)}]{Matsumoto2021}
{Matsumoto}, T. \& {Piran}, T. 2021, MNRAS, 507, 4196

\bibitem[{{Matsumoto} \& {Piran}(2023)}]{Matsumoto2023}
{Matsumoto}, T. \& {Piran}, T. 2023, MNRAS, 522, 4565

\bibitem[{{Matsumoto} \& {Piran}(2024)}]{Matsumoto2024}
{Matsumoto}, T. \& {Piran}, T. 2024, ApJ, 971, 49

\bibitem[{{Metzger} \& {Stone}(2016)}]{Metzger2016}
{Metzger}, B.~D. \& {Stone}, N.~C. 2016, MNRAS, 461, 948

\bibitem[{{Mignone} {et~al.}(2007){Mignone}, {Bodo}, {Massaglia}, {Matsakos},
  {Tesileanu}, {Zanni}, \& {Ferrari}}]{Mignone2007}
{Mignone}, A., {Bodo}, G., {Massaglia}, S., {et~al.} 2007, ApJS, 170, 228

\bibitem[{{Mizumoto} {et~al.}(2021){Mizumoto}, {Nomura}, {Done}, {Ohsuga}, \&
  {Odaka}}]{Mizumoto2021}
{Mizumoto}, M., {Nomura}, M., {Done}, C., {Ohsuga}, K., \& {Odaka}, H. 2021,
  MNRAS, 503, 1442

\bibitem[{{Mou} {et~al.}(2022){Mou}, {Wang}, {Wang}, \& {Yang}}]{Mou2022}
{Mou}, G., {Wang}, T., {Wang}, W., \& {Yang}, J. 2022, MNRAS, 510, 3650

\bibitem[{{Murray} \& {Chiang}(1995)}]{Murray1995}
{Murray}, N. \& {Chiang}, J. 1995, ApJL, 454, L105

\bibitem[{{Nomura} \& {Ohsuga}(2017)}]{Nomura2017}
{Nomura}, M. \& {Ohsuga}, K. 2017, MNRAS, 465, 2873

\bibitem[{{Nomura} {et~al.}(2016){Nomura}, {Ohsuga}, {Takahashi}, {Wada}, \&
  {Yoshida}}]{Nomura2016}
{Nomura}, M., {Ohsuga}, K., {Takahashi}, H.~R., {Wada}, K., \& {Yoshida}, T.
  2016, PASJ, 68, 16

\bibitem[{{Ohsuga} {et~al.}(2005){Ohsuga}, {Mori}, {Nakamoto}, \&
  {Mineshige}}]{Ohsuga2005}
{Ohsuga}, K., {Mori}, M., {Nakamoto}, T., \& {Mineshige}, S. 2005, ApJ, 628,
  368

\bibitem[{{Owocki} {et~al.}(1988){Owocki}, {Castor}, \& {Rybicki}}]{Owocki1988}
{Owocki}, S.~P., {Castor}, J.~I., \& {Rybicki}, G.~B. 1988, ApJ, 335, 914

\bibitem[{{Park} \& {Hayasaki}(2020)}]{Park2020}
{Park}, G. \& {Hayasaki}, K. 2020, ApJ, 900, 3

\bibitem[{{Parkinson} {et~al.}(2022){Parkinson}, {Kingge}, {Matthews}, {Long},
  {Higginbottom}, {Sim}, \& {Mangham}}]{Parkinson2022}
{Parkinson}, E.~J., {Kingge}, C., {Matthews}, J.~H., {et~al.} 2022, MNRAS, 510,
  5426

\bibitem[{{Perlman} {et~al.}(2022){Perlman}, {Meyer}, {Wang}, {Henriksen},
  {Irwin}, {Li}, {Wiegert}, {Li}, \& {Yang}}]{Perlman2022}
{Perlman}, E.~S., {Meyer}, E.~T., {Wang}, Q. D.~{Yuan}, Q., {et~al.} 2022, ApJ,
  925, 143

\bibitem[{{Phinney}(1989)}]{Phinney1989}
{Phinney}, E.~S. 1989, in Morris M., ed., IAU Symp. Vol. 136, Manifestations of
  a Massive Black Hole in the Galactic Center. Kluwer, Dordrecht, p. 543

\bibitem[{{Piran} {et~al.}(2015){Piran}, {Svirski}, {Krolik}, {Cheng}, \&
  {Shiokawa}}]{Piran2015}
{Piran}, T., {Svirski}, G., {Krolik}, J., {Cheng}, R.~M., \& {Shiokawa}, H.
  2015, ApJ, 806, 164

\bibitem[{{Piro} \& {Lu}(2020)}]{Piro2020}
{Piro}, A. \& {Lu}, W. 2020, ApJ, 894, 2

\bibitem[{{Price} {et~al.}(2024){Price}, {Liptai}, {Mandel}, {Shepherd},
  {Lodato}, \& {Levin}}]{Price2024}
{Price}, D.~J., {Liptai}, D., {Mandel}, I., {et~al.} 2024, ApJ, 971, L46

\bibitem[{{Proga} \& {Kallman}(2004)}]{Proga2004}
{Proga}, D. \& {Kallman}, T.~R. 2004, ApJ, 616, 688

\bibitem[{{Proga} {et~al.}(1998){Proga}, {Stone}, \& {Drew}}]{Proga1998}
{Proga}, D., {Stone}, J.~M., \& {Drew}, J.~E. 1998, MNRAS, 295, 595

\bibitem[{{Proga} {et~al.}(2000){Proga}, {Stone}, \& {Kallman}}]{Proga2000}
{Proga}, D., {Stone}, J.~M., \& {Kallman}, T.~R. 2000, ApJ, 543, 686

\bibitem[{{Rees}(1988)}]{Rees1988}
{Rees}, M.~J. 1988, Nature, 333, 523

\bibitem[{{Reis} \& {Miller}(2013)}]{Reis2013}
{Reis}, R.~C. \& {Miller}, J.~M. 2013, ApJ, 769, 7

\bibitem[{{Roth} {et~al.}(2016){Roth}, {Kasen}, \& {Guillochon}}]{Roth2016}
{Roth}, N., {Kasen}, D., \& {Guillochon}, J.~{Ramirez-Ruiz}, E. 2016, ApJ, 827,
  3

\bibitem[{{Rybicki} \& {Hummer}(1978)}]{Rybicki1978}
{Rybicki}, G.~B. \& {Hummer}, D.~G. 1978, ApJ, 219, 654

\bibitem[{{Sfaradi} {et~al.}(2024){Sfaradi}, {Beniamini}, {Horesh}, {Bright},
  {Rhodes}, {Williams}, {Fender}, {Leung}, {Murphy}, \& {Green}}]{Sfaradi2024}
{Sfaradi}, I., {Beniamini}, P., {Horesh}, A.~{Piran}, T., {et~al.} 2024, MNRAS,
  527, 7672

\bibitem[{{Sfaradi} {et~al.}(2022){Sfaradi}, {Horesh}, {Fender}, {Williams},
  {Bright}, \& {Schulze}}]{Sfaradi2022}
{Sfaradi}, I., {Horesh}, A., {Fender}, R.~{Green}, D.~A., {et~al.} 2022, ApJ,
  933, 176

\bibitem[{{Shakura} \& {Sunyaev}(1973)}]{Shakura1973}
{Shakura}, N.~I. \& {Sunyaev}, R.~A. 1973, A\&A, 24, 337

\bibitem[{{Shiokawa} {et~al.}(2015){Shiokawa}, {Krolik}, {Cheng}, {Piran}, \&
  {Noble}}]{Shiokawa2015}
{Shiokawa}, H., {Krolik}, J.~H., {Cheng}, R.~M., {Piran}, T., \& {Noble}, S.~C.
  2015, ApJ, 804, 85

\bibitem[{{Steinberg} \& {Stone}(2024)}]{Steinberg2024}
{Steinberg}, E. \& {Stone}, N.~C. 2024, Nature, 625, 463

\bibitem[{{Strubbe} \& {Quataert}(2009)}]{Strubbe2009}
{Strubbe}, L.~E. \& {Quataert}, E. 2009, MNRAS, 400, 2070

\bibitem[{{Tamilan} {et~al.}(2024){Tamilan}, {Hayasaki}, \&
  {Suzuki}}]{Tamilan2024}
{Tamilan}, M., {Hayasaki}, K., \& {Suzuki}, T.~K. 2024, ApJ, 975, 94

\bibitem[{{Tamilan} {et~al.}(2025{\natexlab{a}}){Tamilan}, {Hayasaki}, \&
  {Suzuki}}]{Tamilan2025a}
{Tamilan}, M., {Hayasaki}, K., \& {Suzuki}, T.~K. 2025{\natexlab{a}}, Progress
  of Theoretical and experimental Physics, b3E02

\bibitem[{{Tamilan} {et~al.}(2025{\natexlab{b}}){Tamilan}, {Hayasaki}, \&
  {Suzuki}}]{Tamilan2025b}
{Tamilan}, M., {Hayasaki}, K., \& {Suzuki}, T.~K. 2025{\natexlab{b}},
  (arXiv:2502.12549)

\bibitem[{{Teboul} \& {Metzger}(2023)}]{Teboul2023}
{Teboul}, O. \& {Metzger}, B.~D. 2023, ApJL, 957, L9

\bibitem[{{Thomsen} {et~al.}(2022){Thomsen}, {Kwan}, {Dai}, \&
  {Ramirez-Ruiz}}]{Thomsen2022}
{Thomsen}, L.~L., {Kwan}, T., {Dai}, L.~{Wu}, S., \& {Ramirez-Ruiz}, E. 2022,
  ApJ, 937, L28

\bibitem[{{Tsuboi} {et~al.}(2016){Tsuboi}, {Kitamura}, {Miyoshi}, {Uehara}, \&
  {Miyazaki}}]{Tsuboi2016}
{Tsuboi}, M., {Kitamura}, Y., {Miyoshi}, M., {Uehara}, K.{Tsutsumi}, T., \&
  {Miyazaki}, A. 2016, PASJ, 68, 7

\bibitem[{{Uno} \& {Maeda}(2020)}]{Uno2020}
{Uno}, K. \& {Maeda}, K. 2020, ApJ, 905, L5

\bibitem[{{Uttley} {et~al.}(2014){Uttley}, {Cackett}, {Fabian}, {Kara}, \&
  {Wilkins}}]{Uttley2014}
{Uttley}, P., {Cackett}, E.~M., {Fabian}, A.~C., {Kara}, E., \& {Wilkins},
  D.~R. 2014, A\&ARv, 22, 72

\bibitem[{{van Velzen} {et~al.}(2020){van Velzen}, {Holoien}, {Hung}, \&
  {Arcavi}}]{Velzen2020}
{van Velzen}, S., {Holoien}, T.~W.~S.~{Onori}, F., {Hung}, T., \& {Arcavi}, I.
  2020, SSRv, 216, 124

\bibitem[{{Wang} {et~al.}(2022){Wang}, {Bu}, \& {Yuan}}]{Wang2022}
{Wang}, W., {Bu}, D., \& {Yuan}, F. 2022, MNRAS, 513, 5818

\bibitem[{{Wevers} {et~al.}(2022){Wevers}, {Nicholl}, {Guolo}, {Gromadzki},
  {Reynolds}, \& {Kankare}}]{Wevers2022}
{Wevers}, T., {Nicholl}, M., {Guolo}, M.~{Charalampopoulos}, P., {et~al.} 2022,
  A\&A, 666, 6

\bibitem[{{Yalinewich} {et~al.}(2019){Yalinewich}, {Steinberg}, {Piran}, \&
  {Krolik}}]{Yalinewich2019}
{Yalinewich}, A., {Steinberg}, E., {Piran}, T., \& {Krolik}, J.~H. 2019, MNRAS,
  487, 4083

\bibitem[{{Yang}(2021)}]{Yang2021a}
{Yang}, X. 2021, ApJ, 922, 262

\bibitem[{{Yang} {et~al.}(2021){Yang}, {Ablimit}, \& {Li}}]{Yang2021b}
{Yang}, X., {Ablimit}, K., \& {Li}, Q. 2021, ApJ, 914, 31

\bibitem[{{Yuan} {et~al.}(2012){Yuan}, {Bu}, \& {Wu}}]{Yuan2012}
{Yuan}, F., {Bu}, D., \& {Wu}, M. 2012, ApJ, 761, 130

\bibitem[{{Zhang} {et~al.}(2024){Zhang}, {Shu}, {Yang}, {Zhang}, {Wang}, {Mou},
  {Zhang}, {Zhou}, \& {Peng}}]{Zhang2024}
{Zhang}, F., {Shu}, X., {Yang}, L.~{Sun}, L., {et~al.} 2024, ApJL, 962, L18

\bibitem[{{Zhao} {et~al.}(2016){Zhao}, {Morris}, \& {Goss}}]{Zhao2016}
{Zhao}, J., {Morris}, M.~R., \& {Goss}, W.~M. 2016, ApJ, 817, L171

\bibitem[{{Zhong} {et~al.}(2023){Zhong}, {Hayasaki}, {Li}, {Berczik}, \&
  {Spurzem}}]{Zhong2023}
{Zhong}, S., {Hayasaki}, K., {Li}, S., {Berczik}, P., \& {Spurzem}, R. 2023,
  ApJ, 959, 19

\bibitem[{{Zhou} {et~al.}(2024){Zhou}, {Zhu}, {Lei}, {Fu}, {Xie}, \&
  {Xu}}]{Zhou2024}
{Zhou}, C., {Zhu}, Z., {Lei}, W., {et~al.} 2024, ApJ, 963, 66

\bibitem[{{Zhuang} {et~al.}(2025){Zhuang}, {Shen}, {Mou}, \& {Lu}}]{Zhuang2025}
{Zhuang}, J., {Shen}, R., {Mou}, G., \& {Lu}, W. 2025, ApJ, 979, 109

\end{thebibliography}

\appendix
\section{Line force multiplier}

We adopt the CAK75 \citep{Castor1975} analytical expression modified by \cite{Owocki1988} to calculate the force multiplier
\begin{equation}
    \mathcal{M}(t) = kt^{-\alpha}\left[\frac{(1+\tau_{\rm max})^{1-\alpha}-1}{\tau_{\rm max}^{1-\alpha}}\right]
    \label{multiplier}
\end{equation}
$t$ is a function of the optical depth parameter defined as
\begin{equation}
    t= \frac{\sigma_e\rho v_{\rm th}}{|dv_l/dl|}
\end{equation}
where $\sigma_e$ is the mass-scattering coefficient for free electrons, $v_{\rm th}$ is the thermal velocity and $dv_l/dl$ is the velocity gradient along the line of sight $\hat{n}$.
In equation (\ref{multiplier}), $k$ is proportional to the total number of lines and calculated as
\begin{equation}
    k = 0.03+0.385\exp{(-1.4\xi^{0.6})}
\end{equation}
where $\xi$ is the ionization parameter defined in Section 2.1. $\alpha = 0.6 $ is the ratio of optically thick to optically thin lines and does not change with $\xi$. $\tau_{\rm max} = t\eta_{\rm max}$ and
\begin{equation}
    \log_{10}\eta_{\rm max} = 6.9\exp{(0.16\xi^{0.4})} \ \ \ \ \ \ \ \ \ \ \ \ \ \ \ \ \ \   {\rm for} \ \  \log_{10} \xi \leq 0.5
\end{equation}
\begin{equation}
    \log_{10}\eta_{\rm max} = 9.1\exp{(-7.96 \times 10^{-3 }\xi)} \ \ \ \ \ \ \  {\rm for} \ \  \log_{10} \xi > 0.5
\end{equation}

The line force also becomes negligible if gas temperature $T>10^5$ K for any value of $\xi$ \citep{Proga2000}.

\section{Derivation of the scaling law for the thin disk density in Equation 8}

The basic equations of the thin disk are as follows \citep{Frank1992},
\begin{equation}
    \overline{\rho} = \Sigma/H
\end{equation}
\begin{equation}
    H=C_{\rm s} r^{3/2}/(GM_{\rm BH})^{1/2}
\end{equation}
\begin{equation}
    C_{\rm s}^2=p/\overline{\rho}
\end{equation}
\begin{equation}
    p=\frac{\overline{\rho}kT_{\rm c}}{\mu m_{\rm p}} + \frac{4\sigma}{3c}T_{\rm c}^4
\end{equation}
\begin{equation}
    \frac{4\sigma T_{\rm c}^4}{3\tau} \approx \frac{3GM_{\rm BH}\dot M}{8\pi r^3}
\end{equation}
\begin{equation}
    \tau=\Sigma \kappa
\end{equation}
\begin{equation}
    \nu\Sigma  \approx \frac{\dot M}{3\pi}
\end{equation}
\begin{equation}
    \nu=\alpha C_{\rm s} H
\end{equation}
In above equations, $\overline{\rho}$ is the density of the disk, $\Sigma$ is the surface density, $H$ is the scale height, $C_{\rm s}$ is the sound speed, $p$ is the pressure including the gas pressure and radiation pressure, $k$ is the Boltzmann constant, $T_{\rm c}$ is the disk temperature, $\sigma$ is the Stefan-Boltzmann constant, $\tau$ is the optical depth, $\kappa$ is the opacity, $\nu$ is the kinematic viscosity and $\alpha$ is the viscosity coefficient.

Substituting Equations (B.1) and (B.8) into Equation (B.7), we have,
\begin{equation}
    \overline{\rho}C_{\rm s}H^2 \propto \dot M
\end{equation}
In the inner region, only considering radiation pressure ($p \propto T_{\rm c}^4$) and using Equation (B.5), we have,
\begin{equation}
    \frac{p}{\tau} = \frac{p}{\Sigma \kappa} = \frac{p}{\overline{\rho} H \kappa} \propto M_{\rm BH}\dot M r^{-3}
\end{equation}
By using Equation (B.3), we have,
\begin{equation}
    C_{\rm s}^2 \propto H M_{\rm BH} \dot M r^{-3}
\end{equation}
Combining Equations (B.2) and (B.11), we have,
\begin{equation}
    H \propto \dot M
\end{equation}
Combing Equations (B.2), (B.9) and (B.12), we have,
\begin{equation}
    \overline{\rho} \propto \dot M^{-2}M_{\rm BH}^{-1/2}r^{3/2}
\end{equation}
Expressing $\dot M$ with the Eddington accretion rate $\dot M = \varepsilon \dot M_{\rm Edd} = \varepsilon \frac{4\pi GM_{\rm BH}m_{\rm p}}{\sigma_{\rm T}c \eta}$, we have,
\begin{equation}
    \overline{\rho} \propto M_{\rm BH}^{-5/2}(r/R_{\rm s})^{3/2}R_{\rm s}^{3/2}(\varepsilon/\eta)^{-2}
\end{equation}
Because $R_{\rm s} \propto M_{\rm BH}$, Equation (B.14) becomes,
\begin{equation}
    \overline{\rho} \propto M_{\rm BH}^{-1}(r/R_{\rm s})^{3/2}(\varepsilon/\eta)^{-2}
\end{equation}
which is the scaling law used in Equation (8).

\section{Dependence on physical parameters}
\subsection{The inner radial boundary}
\begin{figure}[ht]
\includegraphics[width=0.48\textwidth]{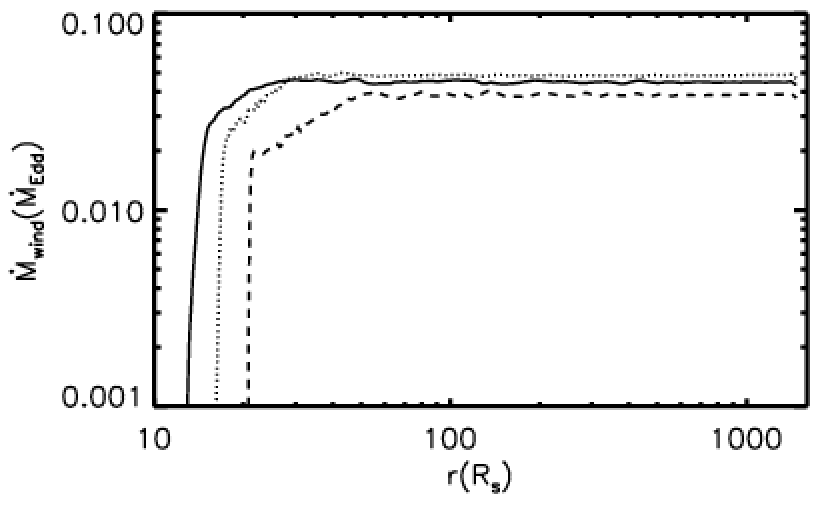}
       \ \centering \caption{Time-averaged radial profiles of wind mass flux for models with inner radial boundary at $10R_{\rm s}$ (solid line, fiducial model in Section 3.1), $15R_{\rm s}$ (dotted line) and $20R_{\rm s}$ (dashed line) with $\varepsilon = 0.8$.}\label{fig:mdotrin}
\end{figure}

We test the effects of radial inner boundary. In this test, we fix the Eddington ratio of disk luminosity to be $\varepsilon=0.8$. In our fiducial model, we have an inner radial boundary of $10R_{\rm s}$. We run two test simulations with inner radial boundary at $15R_{\rm s}$ and $20R_{\rm s}$, respectively. We show the radial profiles of the wind mass flux in Figure \ref{fig:mdotrin}. At the outer radial boundary, the mass flux of wind differs by a factor smaller than $10\%$. Slightly changing the inner radial boundary would not affect the results much.

\subsection{The disk outer boundary}
\begin{figure}[ht]
\includegraphics[width=0.48\textwidth]{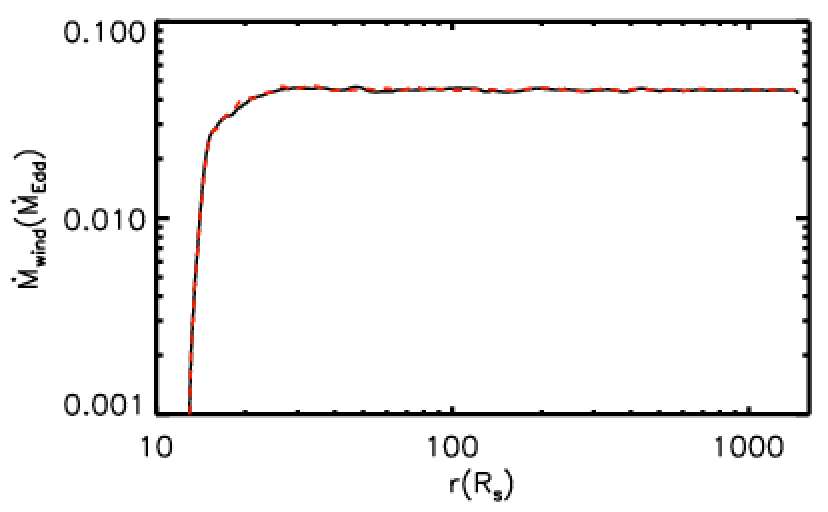}
       \ \centering \caption{Time-averaged radial profile of wind mass flux for models with TDE disk outer boundary at $47R_{\rm s}$ (black solid line, fiducial model in Section 3.1), $100R_{\rm s}$ (red dashed line) with $\varepsilon = 0.8$.}\label{fig:mdotrout}
\end{figure}

We test the effects of TDE disk outer boundary. In this test, we fix the Eddington ratio of disk luminosity to be $\varepsilon=0.8$. In our fiducial model, we have a disk outer boundary of $47R_{\rm s}$. We run a test simulation with the TDE disk outer boundary at $100R_{\rm s}$. We show the radial profiles of the wind mass flux in Figure \ref{fig:mdotrout}. We can see that the radial profiles of the two models are roughly same. Therefore, changing the outer boundary of the TDE disk in reasonable regime would not affect the results.

\subsection{The ratio of X-ray luminosity to disk luminosity}

\begin{figure}[ht]
\includegraphics[width=0.48\textwidth]{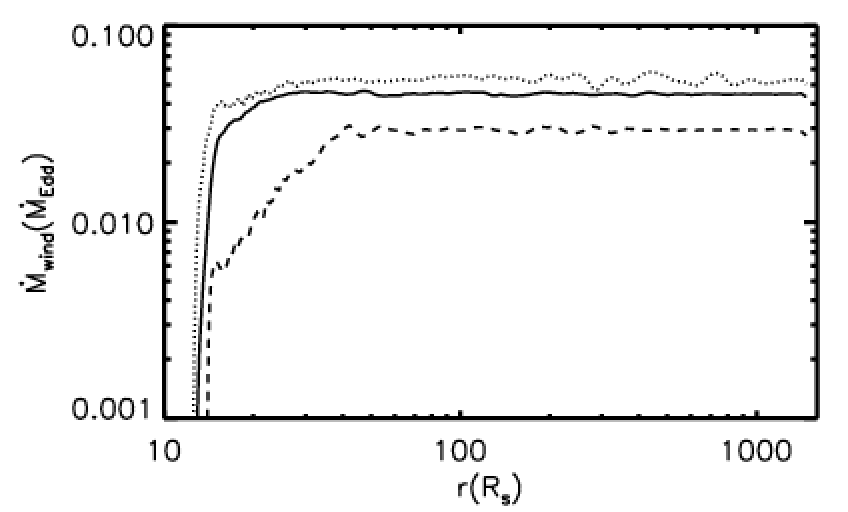}
       \ \centering \caption{Time-averaged radial profile of wind mass flux for models with $f_{\rm X} = 0.03$ (solid line, fiducial model in Section 3.1), $f_{\rm X} = 0.015$ (dotted line) and $f_{\rm X} = 0.05$ (dashed line) with $\varepsilon = 0.8$.}\label{fig:mdotfx}
\end{figure}

We test the effects of the value of $f_{\rm X}$. In this test, we fix the Eddington ratio of disk luminosity to be $\varepsilon=0.8$. In our fiducial model, we have $f_{\rm X} = 0.03$. \cite{Li2019} studied a sample of luminous AGNs. Generally, they found that the ratio of the X-ray luminosity to the bolometric luminosity decreases with the Eddington ratio $\varepsilon$. In their table 2, for the AGNs with Eddington ratios in the range 0.3-1.0, the value of $f_{\rm X}$ is in the range of 0.015-0.05. Therefore, we run two test simulations with $f_{\rm X} = 0.015$ and $f_{\rm X} = 0.05$, respectively. We show the radial profiles of the wind mass flux in Figure \ref{fig:mdotfx}. Generally, the line force multiplier decreases with the increase of $f_{\rm X}$. With fixed UV radiation flux, the line force decreases with the increase of $f_{\rm X}$. Therefore, we can see that the mass flux of wind decreases with the increase of $f_{\rm X}$. However, there is just little change of wind flux. Compared to the fiducial model with $f_{\rm X} = 0.03$, the wind mass flux is changed by a factor smaller than 1.

\section{The Eddington ratio dependence of line force}
\begin{figure}[ht]
\includegraphics[width=0.48\textwidth]{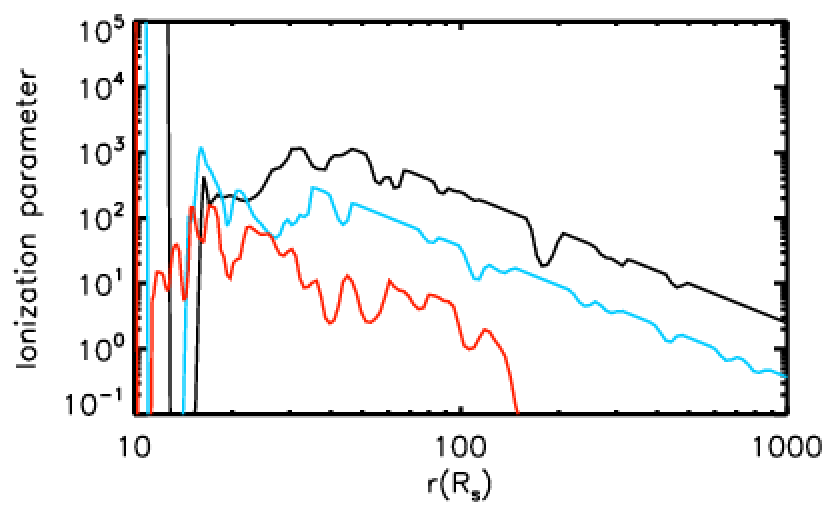}
       \ \centering \caption{Radial profiles of gas ionization parameter at a snapshot when the simulations achieves quasi-steady states. The figure is for an angular position at $\theta=82^\circ$ close to the wind launching mid-plane. The black, blue and red lines correspond to models (introduced in Section 3) of $\varepsilon=0.8$, $\varepsilon=0.6$ and $\varepsilon=0.4$, respectively.}\label{fig:ionpar}
\end{figure}

In section 3.2, we find that the wind is strongest when $\varepsilon = 0.6$. The reason is as follows discussed in Secton 3.2. With the increase of $\varepsilon$, the X-ray flux increases, which will result in high ionization parameter. We show the results of the radial profiles of gas ionization parameter at a snapshot when the simulations achieves quasi-steady states for an angular position at $\theta=82^\circ$ close to the wind launching mid-plane in Figure \ref{fig:ionpar}. It is clear that higher $\varepsilon$ results in higher ionization parameter.

\begin{figure}[ht]
\includegraphics[width=0.48\textwidth]{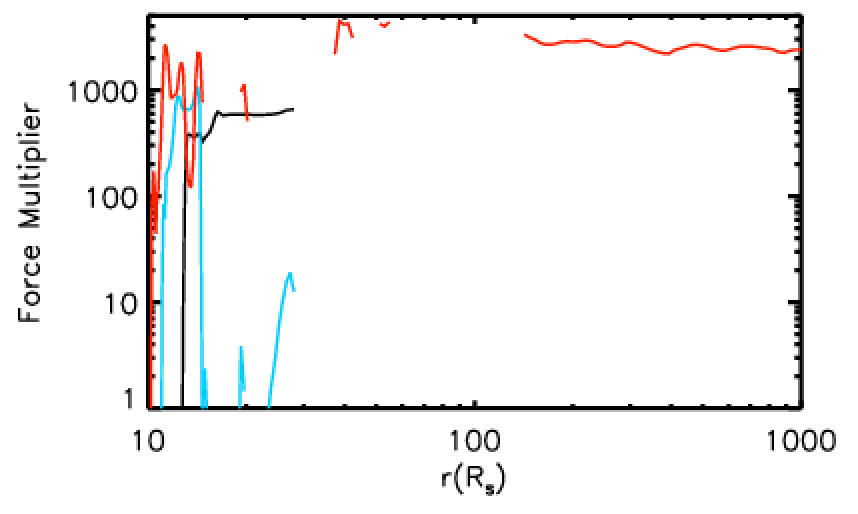}
       \ \centering \caption{Radial profiles of line force multiplier at a snapshot when the simulations achieves quasi-steady states. The figure is for an angular position at $\theta=82^\circ$ close to the wind launching mid-plane. The black, blue and red lines correspond to models (introduced in Section 3) of $\varepsilon=0.8$, $\varepsilon=0.6$ and $\varepsilon=0.4$, respectively.}\label{fig:multiplier}
\end{figure}

We show the corresponding radial profiles of line force multiplier in Figure \ref{fig:multiplier}. The force multiplier is not continuous with radius. At some positions the multiplier is zero. The reason is that as motioned above (in Appendix A), the multiplier is zero when the gas temperature is higher than $10^5$ K \citep{Proga2000}. The region with zero multiplier has gas temperature higher than $10^5$K. We focus on the winds launching and accelerating region ($r< 30 R_{\rm s}$). It can be seen that generally the force multiplier increases with decreasing $\varepsilon$ (or ionization parameter).

\begin{figure}[ht]
\includegraphics[width=0.48\textwidth]{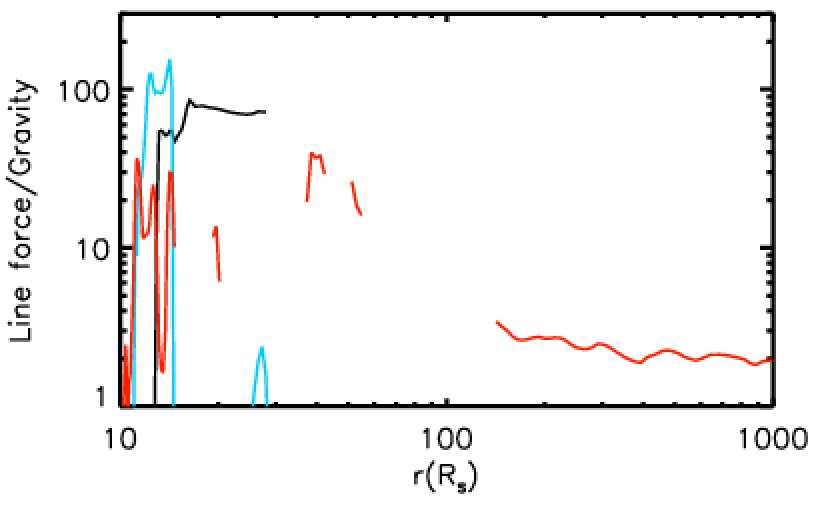}
       \ \centering \caption{Radial profiles of radial component of the line force at a snapshot when the simulations achieves quasi-steady states. The figure is for an angular position at $\theta=82^\circ$ close to the wind launching mid-plane. The black, blue and red lines correspond to models (introduced in Section 3) of $\varepsilon=0.8$, $\varepsilon=0.6$ and $\varepsilon=0.4$, respectively.}\label{fig:lineforce}
\end{figure}

The corresponding radial component of the line forces are shown in Figure \ref{fig:lineforce}. The radial component of line force accelerates wind radially. It can be seen that within the wind launching region $r<30R_{\rm s}$, the line force is strongest in model with $\varepsilon = 0.6$, which has strongest wind.
Therefore, as explained in Section 3.2, the reason for the strongest wind in model with  $\varepsilon = 0.6$ is as follows. With the increase of $\varepsilon$, the ionization parameter of gas increases. The line force multiplier decreases with increasing ionization
parameter. Therefore, when $\varepsilon \geq 0.6$,  the strength of wind decreases with increase of $\varepsilon$. When $\varepsilon < 0.6$, the strength of wind decreases with decreasing of $\varepsilon$. The reason is that smaller $\varepsilon$ results in a smaller radiation flux and radiation pressure.

\begin{figure}[ht]
\includegraphics[width=0.48\textwidth]{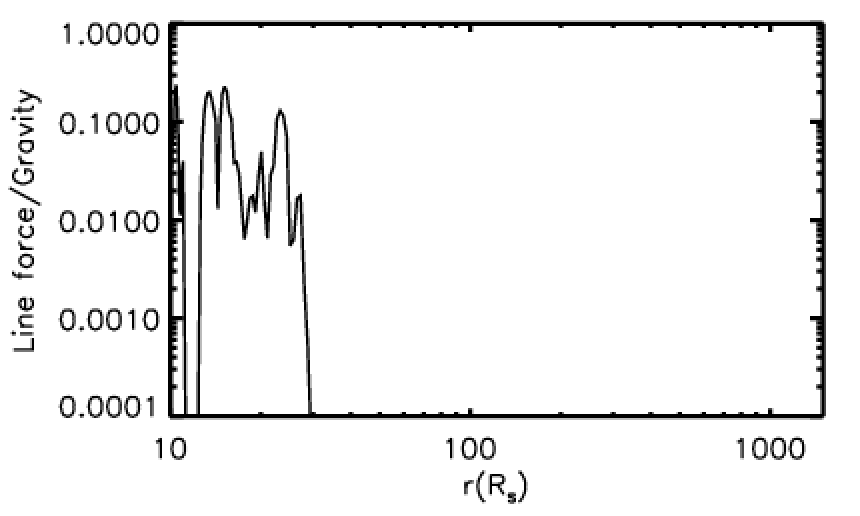}
       \ \centering \caption{Radial profile of radial component of the line force at a snapshot when the simulation achieves quasi-steady states for the model with $\varepsilon=0.2$ (see Section 3). The figure is for an angular position at $\theta=85^\circ$ close to the wind launching mid-plane.}\label{fig:lineforce0.1}
\end{figure}

We plot the radial profile of the radial component of the line force of the model with $\varepsilon=0.2$ (discussed in Section 3) in Figure \ref{fig:lineforce0.1}. The plot is for an angular position of $\theta = 85^\circ$ close to the wind launching mid-plane. It is clear that in this model, the radial line force is much smaller than black hole gravity. Winds can not be driven in this model.

\end{document}